\documentstyle[12pt]{article}

\def\k{k}
\def\kt{\tilde{k}}
\def\gt{\tilde{g}}

\def\Ht{\tilde{H}}

\def\At{\tilde{A}}

\def\St{\tilde{S}}

\begin{document}

\begin{titlepage}

\begin{flushright}
QMW-PH-97-12\\
hep-th/9704021
\end{flushright}

\vspace{3cm}

\begin{center}

{\bf \Large Intrinsic Geometry of D-Branes}

\vspace{.7cm}

M.\ Abou Zeid and C.\ M.\ Hull

\vspace{.7cm}

{\em Physics Department, Queen Mary and Westfield College, \\
Mile End Road, London E1 4NS, U.\ K.\ }

\vspace{.7cm}

and

\vspace{.7cm}

{\em Isaac Newton Institute, 20 Clarkson Road, \\ 
Cambridge CB3 0EH, U.\ K.\ }

\vspace{1cm}

April 1997

\vspace{1cm}

\begin{abstract}

We obtain  forms of Born-Infeld and D-brane actions that are quadratic in derivatives of $X$ and 
linear in $F_{\mu \nu}$ by
  introducing an auxiliary \lq metric' which has both symmetric and anti-symmetric parts,
generalising the 
simplification of the Nambu-Goto action for $p$-branes
using  a symmetric metric. 
The abelian gauge  field appears as a Lagrange multiplier, and solving the constraint
gives the dual form of the $n$ dimensional action 
with an $n-3$ form gauge field instead of a vector gauge field.
We construct the dual action explicitly,  including
  cases which could not be covered previously. The generalisation to supersymmetric D-brane actions
with local fermionic symmetry is also discussed.

\end{abstract}

\end{center}

\end{titlepage}

\section{Actions}                          \label{newactions}

        The Nambu-Goto action for a $p$-brane with $p=n-1$ is
\begin{equation}
S_{NG} = -T_p \int d^n \sigma \sqrt{-\det{ \left( G_{\mu \nu} \right) }}   ,    \label{NG}
\end{equation}
where   $T_p$ is the $p$-brane
tension and
\begin{equation}
G_{\mu \nu} = G_{ij} \partial_\mu X^i \partial_\nu X^j   \label{induced}
\end{equation}
is the world-volume metric induced  by the   spacetime metric $G_{ij}$.
The non-linear form of the action~(\ref{NG}) is inconvenient
for many purposes. However, introducing an intrinsic  worldvolume metric $
g_{\mu
\nu}$ allows one to write down the equivalent action~\cite{Polya,BVH,HT}
\begin{equation}
S_P = -\frac{1}{2}T'_p \int d^n \sigma \sqrt{-g} \left[ g^{\mu \nu} G_{\mu \nu} -(n-2)
\Lambda \right] ,
\label{Polyakov}
\end{equation}
where   $g \equiv \det{(g_{\mu
\nu})}$ and $\Lambda$ is a   constant. The metric $g_{\mu \nu}$ is an auxiliary
field which can be eliminated using its equation of motion
to recover  action~(\ref{NG}). The  constants $T_p$ 
and $T'_{p}$ are related by 
\begin{equation}
T'_p =    \Lambda^{\frac{n}{2}-1} T_p .
\label{tensions}
\end{equation}
This form of the action is much more convenient for many purposes, as it is
quadratic in $\partial X$.

        The Born-Infeld action for a
 vector field $A_\mu$ in
an $n$-dimensional space-time with metric $G_{\mu\nu}$ is
\begin{equation}
S_{BI} = -T_p \int d^n \sigma \sqrt{-\det{\left( G_{\mu \nu} +F_{\mu \nu}
\right) }} ,
\label{DBI}
\end{equation}
where   $F=dA$ is the Maxwell field
strength.
A related $(n-1)$-brane action is
 \begin{equation}
S_{DBI} = -T_p \int d^n \sigma \sqrt{-\det{\left( G_{\mu \nu} +
\cal{F}_{\mu \nu} \right) }} ,
\label{GDBI}
\end{equation}
where $G_{\mu \nu}$ is the induced metric~(\ref{induced}) and
${\cal F}_{\mu \nu}$ is the antisymmetric tensor field
\begin{equation}
{\cal{F}}_{\mu \nu} \equiv F_{\mu \nu} -B_{\mu \nu}      \label{defF}
\end{equation}
with $B_{\mu\nu} $ the pull-back of a space-time 2-form gauge field $B$,
\begin{equation}
B_{\mu \nu} = B_{ij} \partial_\mu X^i \partial_\nu X^j   \label{defB} .
\end{equation}
The action~(\ref{GDBI}) is closely related to the D-brane action, which has been the
subject of much recent work~\cite{CS,AS,Doug,PKT,Ark,Ceral,JSetal1,JSetal2,GHT,BT}
 and the Born-Infeld action~(\ref{DBI}) can be thought of
as a special case of this, but with a different interpretation of $G_{\mu\nu}$.
However, just as in the case of the action~(\ref{DBI}), the non-linearity 
of~(\ref{GDBI})  makes
it rather difficult to study. In particular, dualising the action~(\ref{GDBI}) 
has proved
rather difficult in this approach, and has only been achieved for 
$n \le 5$~\cite{PKT,Ark,JSdual}. Clearly, an action analogous to~(\ref{Polyakov}) 
for this case would be very useful, and it
is the aim of this paper to propose and study just such an action.

The key is to introduce a \lq non-symmetric metric', in the form of  an auxiliary
world-volume tensor
  field
\begin{equation}
\k_{\mu \nu} \equiv g_{\mu \nu} +b_{\mu \nu}         \label{defk}
\end{equation}
with both a symmetric part $g_{\mu \nu}$  and an antisymmetric part $b_{\mu
\nu}$.\footnote{Such \lq metrics' have been used in alternative theories of gravitation;
see e.g.~\cite{AE,Moff}.}  The action which is classically equivalent to
(\ref{GDBI}) is
\begin{equation}
S = - \frac{1}{2}T'_p \int d^n \sigma \sqrt{-\k} \left[ ( {\k}^{-1})^{\mu \nu } \left( G_{\mu \nu} +
\cal{F}_{\mu \nu} \right)
-(n-2)\Lambda \right] ,
\label{EH}
\end{equation}
where $k \equiv \det{\left( k_{\mu \nu}\right)}$; the inverse metric 
$({\k}^{-1})^{\mu \nu}$ satisfies
\begin{equation}
({\k}^{-1} )^{\mu \nu} \k_{\nu \rho} = \delta^{\mu}{}_{\rho} \label{defkinv} .
\end{equation}
Such an action was proposed for Born-Infeld theory in~\cite{UL}. 
For $n \ne 2$, the    $k_{\mu \nu}$ field equation implies   
\begin{equation}
  G_{\mu \nu} +
{\cal{F}}_{\mu \nu} = \Lambda \k_{\nu \mu}       \label{keom}
\end{equation}
and substituting  back into~(\ref{EH}) yields the Born-Infeld-type action~(\ref{GDBI})
where the constants $T_p,T'_p$ are related as in eq.~(\ref{tensions}). 
For $n=2$, the action  (\ref{EH}) is invariant under the generalised Weyl
transformation
\begin{equation}
k_{\mu\nu} \to \omega (\sigma) k_{\mu\nu} 
\label{symm}
\end{equation}
and the $k_{\mu \nu}$ field equation implies   
\begin{equation}
 G_{\mu \nu} + {\cal F}_{\mu \nu}  =  \Omega k_{\nu \mu}       \label{keom2}
\end{equation}
for some conformal factor $\Omega$.

The action~(\ref{EH}) is {\em linear} in $ \left( G_{\mu \nu} +
\cal{F}_{\mu \nu} \right)
$ and so is much easier to analyse
than (\ref{GDBI}).
In particular, it is linear in $F$, so that
$A_\mu$ is a Lagrange multiplier imposing the constraint
\begin{equation}
\partial_\mu \left( \sqrt{-\k} (\k^{-1})^{[\mu \nu]}   \right)  = 0 .
\label{constr}
\end{equation}
The general solution of this is
  $\sqrt{-\k} (\k^{-1})^{[\mu \nu]} = \Ht^{\mu \nu}$, where
\begin{equation}
\tilde{H}^{\mu \nu} \equiv \frac{1}{(n-2)!} \epsilon^{\mu \nu \rho
\gamma_1 \ldots
\gamma_{n-3}}\partial_{[ \rho }\tilde{A}_{\gamma_1 \ldots \gamma_{n-3}]} ,
\label{solH}
\end{equation}
 $\tilde A $ is an   $n-3$ form and $\epsilon^{\mu \nu \rho...}$ 
is the alternating tensor density.
The anti-symmetric part of $k_{\mu\nu}$ can then in principle be solved for in
terms of
$\tilde A$, leaving a dual form of the action involving only the symmetric part
of $k_{\mu\nu}$ and the dual potential $\tilde A$.
To do this explicitly requires a judicious choice of variables, as we now show.

\section{Dual Actions}                               \label{dual}

Instead of introducing a tensor $k_{\mu\nu}$, we introduce a tensor density
$\tilde k^{\mu\nu}$ with $\kt \equiv  \det{( \kt^{\mu \nu} ) }$.
For $n \ne 2$, the action
\begin{equation}
\tilde{S} = -\frac{1}{2} T'_p \int d^n \sigma \left[ \kt^{\mu \nu} \left( G_{\mu \nu}-B_{\mu \nu} + F_{\mu \nu}
\right) -(n-2) (-\kt )^{\frac{1}{n-2}} \Lambda \right]
\label{EHdens}
\end{equation}
is equivalent to~(\ref{EH}), as can be seen by
defining a tensor $k_{\mu\nu}$ by
$({\k}^{-1})^{\mu \nu } = (-\kt )^{-\frac{1}{n-2}} {\kt}^{\mu \nu}$, so that
\begin{equation}
\kt^{\mu \nu } \equiv \sqrt{-\k} (\k^{-1})^{\mu \nu} .
\label{rel1}
\end{equation}
Integrating out $\kt^{\mu \nu}$ yields the action~(\ref{GDBI}) as before.

        Integrating out the world-volume vector field $A_\mu$ from~(\ref{EHdens}) 
gives $\partial_\mu \kt^{[\mu \nu] } = 0$ which is solved by $\kt^{[\mu \nu] }=\Ht^{\mu \nu}$
where $\Ht^{\mu \nu}$ is given in terms of an unconstrained $n-3$ form $\tilde A$ by~(\ref{solH}),
so that 
\begin{equation}
\kt^{\mu \nu} = \gt^{\mu \nu} + \Ht^{\mu \nu} ,
\end{equation}
where the  symmetric tensor density $\gt^{\mu \nu}$ is defined by $\gt^{\mu \nu}\equiv \kt^{(\mu
\nu)}$. 
The action~(\ref{EHdens}) then becomes
\begin{equation}
{S'} = -\frac{1}{2} T'_p \int d^n \sigma \left[ \left(\gt^{\mu \nu} + \Ht^{\mu \nu}
\right)
\left( G_{\mu \nu}-B_{\mu
\nu}  
\right) -(n-2) \Lambda \left(-\det 
{[\gt^{\mu \nu} + \Ht^{\mu \nu}]} \right)^{\frac{1}{n-2}}  \right] .
\label{EHdensdu}
\end{equation}
This is a dual form of the action in which $A_\mu$ has been replaced by $\tilde A$.
It contains the auxiliary symmetric tensor density $\tilde g^{\mu \nu}$ which can in principle be
integrated out; this can be done explicitly for low values of $n$, but is harder for general $n$.¤

We define a symmetric metric tensor $ g_{\mu \nu}$ with inverse $ g^{\mu \nu}$ by
$g^{\mu \nu} = (-\tilde g)^{-\frac{1}{n-2}} \gt^{\mu \nu} $
where $\tilde{g}=\det{(\tilde{g}^{\mu \nu})}$,
so that
\begin{equation}
\gt^{\mu \nu} = \sqrt{-g} g^{\mu \nu} ,
\end{equation}
where  $g=\det{(g_{\mu \nu})}$, and an anti-symmetric tensor by
\begin{equation}
H^{\mu \nu} = {1 \over \sqrt{-g} }\Ht ^{\mu \nu} ,
\end{equation}
so that
 \begin{equation}
\kt^{\mu \nu} = \sqrt{-g} \left( g^{\mu \nu} +H^{\mu \nu} \right)
\end{equation}
and
\begin{equation}
 \kt \equiv \det{(\kt^{\mu\nu})}= -(-g)^{\frac{n}{2}-1} \Delta ,
\end{equation}
where
\begin{equation}
\Delta (g, H) \equiv \det{\left( \delta_{\mu} {}^\nu + H_{\mu} {}^\nu \right)}
\end{equation}
and $H_{\mu}{}^{\nu} = g_{\mu \rho}H^{\rho \nu}$. Then the
action~(\ref{EHdensdu}) can be rewritten as
\begin{equation}
\tilde{S}_D = -\frac{1}{2}T'_p  \int d^n \sigma \sqrt{-g} \left( g^{\mu \nu} G_{\mu \nu} +
\Sigma \right) ,
\label{dualEHdens}
\end{equation}
where
\begin{equation}
\Sigma \equiv -(n-2) \Lambda \Delta ^{\frac{1}{n-2}} -H^{\mu \nu} B_{\mu \nu} .
\end{equation}
The action~(\ref{dualEHdens})  is the dual form of action~(\ref{EHdens}).
Unfortunately, the metric dependence of $\Delta$ makes it hard to eliminate
$g_{\mu \nu}$ from this action explicitly. 

        For $n=2$, the action~(\ref{EH}) has the Weyl symmetry~(\ref{symm})
and can be rewritten using a tensor density $\tilde k^{\mu \nu}$ as
\begin{equation}
\St^2 = -\frac{1}{2} T_1 \int d^2 \sigma \left\{ \kt^{\mu \nu} \left( G_{\mu \nu} + 
{\cal F}_{\mu \nu}
\right) +\lambda \left[ \det{(\kt^{\mu \nu} )} +1 \right]
\right\} .
\end{equation}
Integrating out $\lambda$ yields the constraint
\begin{equation}
\kt =-1
\end{equation}
which is solved in $n=2$ dimensions by
\begin{equation}
{\kt}^{\mu \nu } \equiv \sqrt{-\k} (\k^{-1})^{\mu \nu} ,
\label{solk2}
\end{equation}
so that  one recovers the original
action~(\ref{EH}). If instead one keeps the Lagrange multiplier and
integrates out the world-volume vector $A$, one
finds the constraint eq.~(\ref{constr}) again. For $n=2$, this is solved by
\begin{equation}
{\kt}^{[\mu \nu] } = \epsilon^{\mu \nu}  \Lambda ,
\end{equation}
where $\Lambda$ is a constant.
The dual action for $n=2$ is then
\begin{equation}
\St^2_D = -\frac{1}{2}T_1 \int d^2 \sigma \left\{ \left[ \gt^{\mu \nu}+\Lambda \epsilon^{\mu \nu} \right]  \left[ G_{\mu \nu}  -B_{\mu \nu} \right]
  +\lambda \left[ \det{ (\gt^{\mu \nu}) }  +1+
\Lambda^2
\right]
\right\}
\end{equation}
where $\gt^{\mu \nu}=\kt^{(\mu \nu)}$
and we have used the identity
\begin{equation}
  \det{\left( \gt^{\mu \nu} + \epsilon^{\mu \nu}  \Lambda
\right) }=\det{\left( \gt^{\mu \nu} 
\right) }+\Lambda^2 .
\end{equation}
Integrating out $\lambda $ gives
$\det{ (\gt^{\mu \nu})} =-1-\Lambda^2$, which is solved in terms of an unconstrained metric $g_{\mu
\nu}$ by
\begin{equation}
\gt^{\mu \nu} =\sqrt{1+\Lambda^2}\sqrt{-g} g^{\mu \nu}  
\end{equation}
so that the action becomes
\begin{equation}
S^2_D = -\frac{1}{2}T_1 \int d^2 \sigma \left(\sqrt{1+\Lambda^2} \sqrt{-g} g^{\mu \nu} 
 G_{\mu \nu} +\Lambda\epsilon^{\mu \nu} B_{\mu \nu} \right) .
\label{dual2}
\end{equation}
The metric can be eliminated from this to give the dual action of ref.~\cite{CS,AS,Ark}
\begin{equation}
S^2_D = -T_1 \int d^2 \sigma \left( \sqrt{1+\Lambda^2} \sqrt{-\det{(G_{\mu\nu}
)}}  
  +\frac{1}{2} \Lambda \epsilon^{\mu \nu} B_{\mu \nu} \right) .
\label{dualark}
\end{equation}

\section{More Dual Actions}                               \label{moredual}

Consider actions given by the sum of~(\ref{GDBI}) and some action $S_F=\int d^n\sigma 
f(F)$ which is
algebraic in $F$; in the next section we will be interested in the example of D-brane actions which
are of this form. Defining
\begin{equation}
N_{\mu \nu} \equiv G_{\mu \nu }-B_{\mu \nu}  ,        \label{defN}
\end{equation}
the action can be rewritten in first order form as
\begin{equation}
-\frac{1}{2}T'_p \int d^n \sigma \left\{ \sqrt{-\k} \left[ (\k^{-1})^{\mu \nu} (N_{\mu \nu}+F_{\mu \nu } )
 -(n-2) \Lambda \right] +\frac{1}{2} \tilde{H}^{\mu \nu}
\left( F_{\mu \nu} -2\partial_{[ \mu}A_{\nu ]} \right) +f(F)\right\} .
\label{actf}
\end{equation}
Here the anti-symmetric tensor density $ \tilde{H}^{\mu \nu}$ is a Lagrange multiplier imposing
$F=dA$ and can be   integrated out to regain the original action. Alternatively, integrating over
$A_\mu$ imposes
\begin{equation}
\partial_\mu \tilde{H}^{\mu \nu }=0  ,   \label{constrH}
\end{equation}
which can be solved in terms of an $n-3$ form $\tilde A$ as before:
\begin{equation}
\tilde{H}^{\mu \nu} = \frac{1}{(n-2)!} \epsilon^{\mu \nu \rho \gamma_1 \ldots
\gamma_{n-3}}\partial_{[ \rho }\tilde{A}_{\gamma_1 \ldots \gamma_{n-3}]} .
\label{solHagain}
\end{equation}
Now $F$ is an auxiliary  2-form occuring  algebraically; we emphasize this by 
rewriting $F \to L$ so that the action is
\begin{equation}
-\frac{1}{2}T'_p \int d^n \sigma \left\{ \sqrt{-\k} \left[ (\k^{-1} )^{\mu \nu} \left( N_{\mu \nu}
+L_{\mu \nu} \right) -(n-2) \Lambda \right] +\frac{1}{2}\tilde{H}^{\mu \nu}
L_{\mu \nu} +f(L)\right\} .
\label{EHL}
\end{equation}
The field equation for $L_{\mu \nu}$ is
\begin{equation}
\sqrt{-\k} (\k^{-1})^{[\mu \nu ]} +\frac{1}{2} \tilde{H}^{\mu \nu} +{\delta f\over \delta L_{\mu
\nu}}= 0 .
\label{eomL}
\end{equation}
If $f=0$, this can be used to recover  the dual action~(\ref{dualEHdens}) of the last section. More generally, if
$f(L) $ is at most quartic in $L$, this can be solved to give an expression for $L_{\mu \nu}$
which can then be re-substituted in~(\ref{EHL}) to give a dual action analogous to~(\ref{dualEHdens}). This is  
applicable to the D-brane actions considered in the next two sections, in which $f$ is at most
quartic for $p<9$ branes.

Integrating out $k_{\mu \nu}$ from~(\ref{EHL}) gives
\begin{equation}
 -T_p \int d^n \sigma \left\{ \sqrt{-\det{\left( N_{\mu \nu}+L_{\mu \nu}\right)}}
+\frac{1}{2}\tilde{H}^{\mu \nu}L_{\mu \nu} +{T'_p \over T_p}f(L)\right\} .
\end{equation}
If $f=0$ and $n \le 5$, the equation of motion for $L$ can be solved explicitly and
the solution  substituted in~(\ref{EHL}) to get the dual action~\cite{Ark,JSdual}
\begin{equation}
S_D = -T_p \int d^n \sigma \left\{ \sqrt{-\det{\left( G_{\mu \nu} +iK_{\mu \nu}
\right) }} +\frac{1}{2}\tilde{H}^{\mu \nu}B_{\mu \nu } \right\} ,
\label{BID}
\end{equation}
where
\begin{equation}
K_{\mu \nu} \equiv \frac{1}{\sqrt{-\det{(G_{\mu \nu})}}} G_{\mu \rho}
G_{\nu \lambda} \tilde{H}^{\rho \lambda} .
\label{defK}
\end{equation}
This can in turn be linearised to give the equivalent action
\begin{equation}
-\frac{1}{2}T'_p \int d^n \sigma \left\{ \sqrt{-\k} \left[ (\k^{-1} )^{\mu \nu} \left( G_{\mu \nu}
+iK_{\mu \nu} \right) -(n-2) \Lambda \right] +\frac{1}{2}\tilde{H}^{\mu \nu}
B_{\mu \nu}
\right\} .
\label{EHLa}
\end{equation}

\section{ D-Brane Actions }

\label{Dbrane}

The bosonic part of the
effective world-volume action for a D-brane in a type II supergravity background is
~\cite{CS,Doug,GHT}
\begin{equation}
S_{1} = -T_p \int d^n \sigma e^{-\phi} \sqrt{-\det{\left( G_{\mu \nu} +
{\cal F}_{\mu \nu}
\right)}} +T_p \int_{W_{n}} C e^{\cal F} ,
\label{BI+WZ}
\end{equation}
The first term is of the form~(\ref{GDBI}) with an extra dependence on
the  dilaton
field $\phi$. The second term is a Wess-Zumino term and gives the coupling to the
background Ramond-Ramond
$r$-form gauge fields $C^{(r)}$ (where $r$ is odd for type IIA and even for type IIB). The
potentials $C^{(r)}$ for $r>4$ are the duals of the potentials $C^{(8-r)}$.
 In~(\ref{BI+WZ}),
$C$ is the formal sum~\cite{GHT}
\begin{equation}
C \equiv \sum_{r=0}^{9} C^{(r)}  ,       \label{defC}
\end{equation}
all forms in space-time are pulled back to the worldvolume of the brane
$W_n$ and it is understood that the $n$-form part of $C e^{\cal F}$, which is $C^{(n)} +C^{(n-2)}
{\cal F} +\frac{1}{2} C^{(n-4)} {\cal F}^2 +\ldots $, is selected.
The case of the 9-form potential $C^{(9)}$ is special because its equation of
motion implies that the dual of its field strength is a constant $m$. This
constant will be taken to be zero here, so that $C^{(9)}=0$; the more general situation
will be discussed elsewhere~\cite{AH}.  

Introducing $k_{\mu \nu}$,     we obtain the classically equivalent     
 D-brane action 
\begin{equation}
S'_{1}= -\frac{1}{2}T'_p \int d^n \sigma  \sqrt{-\k} e^{-\phi } \left[ (\k^{-1} )^{\mu
\nu} \left( G_{\mu \nu} +{\cal F}_{\mu \nu} \right) - (n-2) \Lambda \right]
+ T_p \int_{W_{n}} Ce^{\cal F} .
\label{P+WZ}
\end{equation}
 The field 
equation for $k_{\mu \nu}$ is given in~(\ref{keom}); substituting
back into~(\ref{P+WZ}) yields~(\ref{BI+WZ}).
The action is of the form~(\ref{actf}) (apart from the introduction of the dilaton) 
so that the dual action is
(cf.~(\ref{EHL}))
\begin{eqnarray}
& & -\frac{1}{2}T'_p \int d^n \sigma \left\{ \sqrt{-\k}e^{-\phi} \left[ (\k^{-1} )^{\mu \nu} \left( N_{\mu \nu}
+L_{\mu \nu} \right) -(n-2) \Lambda \right] +\frac{1}{2}\tilde{H}^{\mu \nu}
L_{\mu \nu}  \right\}  \nonumber \\ & & + T_p \int_{W_{n}} Ce^{L-B}.
\label{EHLD}
\end{eqnarray}
The potential $f(L)\sim Ce^{L-B}$ is a polynomial
of order $[n/2]$ in $L$ (i.e. the integer part of $n/2$),
so that the
  field equation for $L_{\mu \nu}$~(\ref{eomL}) is of order $[n/2]-1$ in $L$ and so should be soluble
explicitly for all $n \le 10$. 
In particular, it is quadratic for $n\le 8$, so that the dual action for p-branes with $p\le 7$ can 
be obtained straightforwardly.
This will be discussed elsewhere~\cite{AH}; here we will
consider only the case in which
 $C^{(n-4)} =C^{(n-6)}=C^{(n-8)}=0$ so that the action is linear in $F$.        Then $A$ is a
Lagrange multiplier imposing the  constraint
\begin{equation}
\partial_\mu \left( \sqrt{-k} (k^{-1})^{[ \mu \nu ]} e^{-\phi}
-\frac{2 T_p}{T'_p} \frac{1}{(n-2)!} \epsilon^{\mu \nu \gamma_1 \ldots \gamma_{n-2}}
(C^{(n-2)})_{\gamma_1 \ldots \gamma_{n-2}} \right) = 0 .
\label{constrWZ}
\end{equation}
The general solution of this constraint is
\begin{equation}
\sqrt{-k} (k^{-1} )^{[ \mu \nu ]} e^{-\phi} = \Ht^{\mu \nu} 
+\frac{2 T_p}{T'_p} \frac{1}{(n-2)!} \epsilon^{\mu \nu \gamma_1 \ldots \gamma_{n-2}}
(C^{(n-2)})_{\gamma_1 \ldots \gamma_{n-2}} \equiv \tilde{{\cal H}}^{\mu \nu} ,
\label{defHtilde}
\end{equation}
where $\Ht^{\mu \nu}$ is  given in terms of $\At$ by~(\ref{solHagain}).

        To obtain the dual action, we first express~(\ref{P+WZ})
in terms of a density ${\kt}^{\mu \nu}$. For $n \ne 2$, this gives the equivalent
action
\begin{eqnarray}
\St_1 & = & -\frac{1}{2} T'_p \int d^n \sigma e^{-\phi} \left[ 
{\kt}^{\mu \nu} \left( G_{\mu \nu} +{\cal F}_{\mu \nu} \right) 
-(n-2) (-{\kt})^{\frac{1}{n-2}}\Lambda \right] \nonumber \\
& & +T_p \int_{W_{n}} C^{(n)}+ C^{(n-2)}{\cal F} .
\label{S1tdens}
\end{eqnarray}
 Integrating out ${\kt}^{\mu
\nu}$ yields the action~(\ref{BI+WZ}), while integrating out 
$A$ gives 
\begin{equation}
{\kt}^{\mu \nu} = \gt^{\mu \nu} +\tilde{{\cal H}}^{\mu \nu} ,
\end{equation}
where $\tilde{{\cal H}}^{\mu \nu}$ is given in terms of $\At$ and $C^{(n-2)}$ 
by eq.~(\ref{defHtilde}), and $\gt^{\mu \nu}$ is a symmetric tensor 
density. The action~(\ref{S1tdens}) can be written in terms of tensors
$g^{\mu \nu} = (-\gt )^{-\frac{1}{n-2}} \gt^{\mu \nu}$ (with $\gt =\det{(
\gt^{\mu \nu})}$)
and ${\cal H}^{\mu \nu} = (-g)^{-\frac{1}{2}} \tilde{{\cal H}}^{\mu \nu}$
(with $g=\det{(g_{\mu \nu})}$) as
\begin{eqnarray}
\St_{1D} & =  & -\frac{1}{2}T'_p \int d^n \sigma  e^{-\phi}
\sqrt{-g} \left[ \left( g^{\mu \nu} +{\cal H}^{\mu \nu} \right)
N_{\mu \nu} -(n-2) {{\Omega}}^{\frac{1}{n-2}} \Lambda \right] 
\nonumber \\ & & +T_p \int_{W_{n}} C^{(n)}- C^{(n-2)}B,
\label{S1Dl}
\end{eqnarray}
where 
\begin{equation}
{\Omega} \equiv \det{ \left( \delta_{\mu}{}^{\nu} + {\cal H}_{\mu}{}^{\nu}
\right)}
\end{equation}
and ${\cal H}_{\mu}{}^{\nu} = ({\gt}^{-1})_{\mu \rho} {\cal H}^{\rho \nu}$. The
action~(\ref{S1Dl}) is the dual form of action~(\ref{P+WZ}). Again, the
dependence of ${\Omega}$ on the metric $g_{\mu \nu}$ makes the elimination
of the latter from the action difficult, although possible in principle.

        For $n=2$, the action~(\ref{P+WZ}) still possesses the generalised
Weyl symmetry~(\ref{symm}) and can be rewritten in terms of a
tensor density ${\kt}^{\mu \nu}$ as
\begin{equation}
\St_1^2 = -\frac{1}{2}T_1 \int d^2 \sigma e^{-\phi} \left[
{\kt}^{ \mu \nu }  \left( N_{\mu \nu}+ F_{\mu \nu} \right) +\lambda
\left( \kt +1 \right) \right] + T_1 \int_{W_{2}}
C^{(2)}+ C^{(0)}{\cal F} .
\label{P+WZ2}
\end{equation}
Integrating out $\lambda$ yields the constraint ${\kt} = -1$, which is
solved by eq.~(\ref{solk2}), so that one recovers the original action. Keeping
the Lagrange multiplier and integrating out $A$ one finds the constraint
\begin{equation}
\partial_\mu \left( e^{-\phi} {\kt}^{[\mu \nu ]} - 
2\epsilon^{\mu \nu}
C^{(0)} \right)=0 ,
\end{equation} 
which for $n=2$ is solved by 
\begin{equation}
e^{-\phi} {\kt}^{[\mu \nu ]} = \epsilon^{\mu \nu} {\cal  E} ,
\end{equation} 
where 
\begin{equation}
{\cal E} \equiv   \tilde{\Lambda}
+ 2C^{(0)}   
\end{equation}
and $\tilde{\Lambda}$ is a constant.
The dual action for $n=2$ is then
\begin{eqnarray}
\St_{1D}^2 & = & -\frac{1}{2}T_1 \int d^2 \sigma e^{-\phi} \left[ \left( \gt^{\mu \nu} +
e^{\phi} \epsilon^{\mu \nu} {\cal E} \right) N_{\mu \nu}+
\lambda \left( \det{( \gt^{\mu \nu})}  +1 +e^{2\phi}  {\cal E}^{2}\right) \right] \nonumber \\ & & +T_1 \int_{W_{2}} C^{(2)}- C^{(0)}{B} ,
\label{S1Dl2}
\end{eqnarray}
where $\gt^{\mu \nu} = \kt^{(\mu \nu )}$. 
Integrating out $\lambda$ gives   the
dual action in the form
\begin{equation}
S_1^2 = -\frac{1}{2}T_1 \int d^2 \sigma    
\sqrt{e^{-2\phi}+{\cal E}^2}
 \sqrt{-g}g^{\mu \nu}
  G_{\mu \nu}     +T_1 \int_{W_{2}} 
C^{(2)}+({\cal E}- C^{(0)}){B} .
\end{equation}
Finally, integrating out $g_{\mu \nu}$ gives the action~\cite{Ark}
        \begin{equation}
S_1^2 = -T_1 \int d^2 \sigma   
\sqrt{e^{-2\phi}+{\cal E}^2}
 \sqrt{- \det{\left(
  G_{\mu \nu}\right) }}   +T_1 \int_{W_{2}} 
C^{(2)}+({\cal E}- C^{(0)}){B} .
\end{equation}

\section{Supersymmetric D-Brane Actions}

        The new actions discussed above  can be extended to supersymmetric
D-brane actions with local kappa symmetry equivalent to those presented
in refs.~\cite{Ceral,BT,JSetal1,JSetal2} at the classical level.

        The (flat) superspace coordinates are the $D=10$ space-time coordinates $X^i$
and the Grassmann coordinates $\theta$, which are space-time spinors and
world-volume scalars. For the type IIA superstring (even $p$), $\theta $ is
Majorana but not Weyl while in the IIB superstring there are two
Majorana-Weyl spinors $\theta_\alpha$ ($\alpha = 1,2$) of the same chirality.
The superspace (global) supersymmetry transformations are
\begin{equation}
\delta_\epsilon \theta = \epsilon \ \ , \ \ \delta_\epsilon X^i =
\overline{\epsilon} \Gamma^i \theta .
\label{susytransf}
\end{equation}

        The world-volume theory has global type IIA or type IIB
super-Poincar\'{e} symmetry and is   constructed using the
supersymmetric one-forms $\partial_\mu \theta$ and
\begin{equation}
\Pi^{i}_{\mu} = \partial_\mu X^i -\overline{\theta} \Gamma^i \partial_\mu
\theta .
\label{defpi}
\end{equation}
The induced world-volume metric is
\begin{equation}
G_{\mu \nu} = G_{ij}\Pi^{i}_{\mu} \Pi^{j}_{\nu} .
\label{susyinduced}
\end{equation}
The   supersymmetric   world-volume gauge
field-strength  two form ${\cal F}_{\mu \nu}$ is given by~(\ref{defF}) for the
following choice of the two form $B$~\cite{PKT}
\begin{equation}
B = -\overline{\theta} \Gamma_{11} \Gamma_i d\theta
\left( dX^i +\frac{1}{2} \overline{\theta} \Gamma^i d\theta \right)
\label{paul}
\end{equation}
when $p$ is even or the same formula with $\Gamma_{11}$ replaced with the Pauli
matrix $\tau_3$ when $p$ is odd. With the choice~(\ref{paul}), $\delta_\epsilon
B$ is an exact two-form and $\cal F$ is supersymmetric for an appropriate
choice of $\delta_\epsilon A$~\cite{JSetal2}.

        The effective world-volume action for a D-brane in 
flat superspace with constant dilaton is
\begin{equation}
S^{DBI}_{1} = -T_p \int d^n \sigma e^{-\phi}\sqrt{-\det{ \left( G_{\mu \nu} +
{\cal F}_{\mu \nu} \right)}} +T_p \int_{W_{n}} C e^{\cal F} ,
\label{S1}
\end{equation}
which is formally of the same form as~(\ref{BI+WZ}).
Again $C$ represents a complex of
differential forms $C^{(r)}$, as in~(\ref{defC}), but now the $r$-forms
$C^{(r)}$ are the pull-backs of superspace forms 
$C^{(r)}=d \bar \theta T^{(r-2)} d\theta$ for certain $r-2$ forms $T^{(r-2)}$ given explicitly 
in~\cite{Ceral,BT,JSetal1,JSetal2}.
This action is    supersymmetric and  
invariant under   local
kappa symmetry~\cite{Ceral,BT,JSetal1,JSetal2}.

      A classically equivalent form of the 
 D-brane action is given by   
\begin{equation}
S^{P}_{1} = -\frac{1}{2}T'_p \int d^n \sigma e^{-\phi}\sqrt{-\k} \left[ (\k^{-1} )^{\mu \nu}
\left( G_{\mu \nu} + {\cal F}_{\mu \nu} \right) - (n-2) \Lambda \right] +T_p \int_{W_{n}} C e^{\cal
F} 
\label{SP1}
\end{equation}
  The action is of the same form as~(\ref{P+WZ}) and   can be dualised to 
give~(\ref{EHLD}) for $n\ne 2$ or~(\ref{S1Dl2})
for $n=2$.

\vspace{1.5cm}

{\bf Acknowledgements}
\\
\\
        We would like to thank 
John Schwarz  for discussions. The work of M.\ A.\ is supported in part by
the Overseas Research Scheme and in part by Queen Mary and Westfield College, London.

\end{document}